# MAPPING A CLOUD OF ULTRACOLD ATOMS ONTO A MINIATURE STORAGE RING

Wilbert Rooijakkers*

*Harvard University, Physics Department,*
*Lyman Laboratory, Cambridge MA 02138*

We describe how to realize magnetic and magneto-optical confinement of ultracold atoms in a torus with adjustable diameter and how an elliptical cloud of ultracold atoms can be adiabatically transformed to have a toroidal shape. An experiment with cold $^{87}$Rb atoms demonstrates the feasibility of shape transformations. These techniques can be used for atom interferometry and quantum computation.

PACS #: 03.75.Be, 32.80.Pj, 03.67.Lx

*) present address: University of Melbourne, School of Physics, Victoria 3010, Australia,
e-mail: wilbert@physics.unimelb.edu.au

1 . Introduction

Ultracold atoms in a magnetic holding field provide an ideal system for studying quantum phase coherence [1] and spin coherence [2], due to the nearly perfect isolation from the environment to where quantum information may escape. Indeed, these atoms can be prepared in distinct quantum states of the holding potential, a single ground state in the case of bosons or filled up from the ground state to the Fermi level in the case of fermions. The exploitation of the coherence properties of cold atom clouds has only just started but promises great opportunities in the fields of atom interferometry [3] and quantum information [4].

Many groups are looking to improve the integration and complexity of the magnetic fields that hold the atoms, to create what has become known as 'atom chips', and it has been demonstrated that a Bose Einstein Condensate (BEC) can be manipulated at distances as small as 50 μm from a patterned surface [5,6]. Usually the cold atoms are collected and pre cooled far away from the surface (> 1mm) after which the cloud is continuously and adiabatically transported, while maintaining its spherical shape [7]. We demonstrate in this work that the same process can be used to adiabatically deform the cloud into more complex shapes such as a ring.

Coherent particles confined to a ring are a fundamental system in the study of quantum mechanics, and have been extensively studied in solid-state physics using liquid helium [8] and using electrons in micrometer sized solid-state devices [9]. Atom interferometry in a ring can be used to build an ultra sensitive gyroscope, using the Sagnac phase shift that is incurred for atoms traveling in opposite directions along the ring. Not only is a ring the simplest structure that could facilitate such a rotation sensor; it also is robust to systematic errors due to its inherent symmetry. It has recently been demonstrated that super conducting flux Q-bits can be realized in solid state devices where the system is prepared in a coherent state of oppositely running currents [10], and that these Q-bits can become entangled [11]. Essentially the same physics would apply to cold atoms in a magnetic holding field.

On the fundamental level a new physics regime is entered when atoms are confined to length scale of the order of the elastic scattering length. In this case, sometimes described as the 'Tonks' gas regime, Bose statistics no longer holds due the fact that particles cannot pass by each other. Although this system has been studied theoretically both in straight [12] and ring [13,14] geometries, experimental realization has been elusive. In our novel scheme the question of reaching stronger confinement in a toroidal field is purely answered by the technical ability to increase the current that produces the magnetic field, without affecting the longitudinal degree of freedom. Thus the atoms will eventually enter the collisional regime and subsequently the Tonks regime, for sufficient scaling of the magnetic field, without loss of particles.

Storage rings for cold atoms have been proposed [15] and realized [16,17] before but these schemes load atoms to a small section of the ring. To fill an entire ring we propose to confine weak field seeking states of the ground state Zeeman multiplet in a toroidal magnetic field, created by a single hollow cylindrical shield of mu-metal, wound with kapton wire (Figure 1) combined with an external bias field. Ferromagnetic material, such as mu-metal, can be used to concentrate the magnetic field generated by many current carrying wires along the long axis of the shield. The spatial structure of the magnetic field depends on the specific properties, the treatment and the geometry of the shield, but can be described in a very good approximation by a uniformly poled material with constant magnetic field inside. We have demonstrated before that we can generate surface fields of up to 500 Gauss, with relatively small electric power (<100 Watt). Combinations of shields can be used to make confining magnetic guides for cold atoms [18], and very recently we have loaded and transported $^{87}$Rb atoms along a stadium shaped ring with a circumference of 10 cm [19]. In this work we describe a simpler, intrinsically symmetric storage ring that can be scaled down to a circumference of less than 10 mm.

2. Miniature Ring Geometry.

To describe our present system we use capitals X, Y and Z for absolute position coordinates, R for the radius of the hollow coil and D for its length. Dimensionless position coordinates in units of R are represented by small characters, e.g. x=X/R. Magnetic field derivatives are given by using an appropriate number of apostrophes, e.g. B'$_x$=dB/dX. Figure 1 shows the coil positioned such that its symmetry axis coincides with the z-axis and its top coincides with the x-y plane. A thin quarter wave plate, reflection coated at the bottom, is put on top of the coil, and serves to reflect the laser beam that is used to capture atoms in the magnetic field minima above the coil, with the appropriate polarization.

The location of the minima is a function of the applied external bias field, in addition to the field provided by the coil. Instead of integrating current filaments using Biot-Savart's law, we use the equivalence of electric and magnetic dipoles to calculate the field at any specific position. The following analysis is scale invariant and can be applied to ferromagnetic shields of any diameter or shape. An infinitesimal section of the ring shield is considered as a magnetic dipole, or equivalent: an electric dipole with opposite charges, located at Z=0 and Z=-D respectively. The field can now be calculated as the sum of the electric field due to individual charges. The electric potential Φ due to a ring with charge Q can be expressed analytically in closed form [20]:

$$\Phi(\rho, Z) = \frac{Q}{4\pi\varepsilon_0} \frac{2}{\pi} \frac{1}{\sqrt{(\rho+R)^2 + Z^2}} K(k^2) \qquad (1)$$

with $\rho = \sqrt{X^2 + Y^2}$ and K(k$^2$) is the complete elliptic integral of the first kind, and:

$$k^2 = \frac{4\rho R}{(\rho+R)^2 + Z^2} \qquad (2)$$

If the coil is long (D>>X,Y,Z) or close to the top section of the shield (X$^2$+Y$^2$~R$^2$ and Z<<D) the potential can be considered due to the top charged ring only. This situation is likely to occur in a typical experimental situation where one wants to exploit the field integrating capacity of mu-metal. Taking the gradient in cylindrical coordinates in expression (1) provides an expression for the radial (axial) component of the field.

We have studied the location of the field minima as a function of the applied external vertical bias field $B_{bias}$, assuming $B_{bias}$ can be vectorially added to the coil field without changing it. This is not unrealistic, as ferromagnetic materials exist with stiffness against depolarization that exceeds 1 Tesla. The validity of this "superposition" principle was also tested in previous experiments [18] where the measured field agreed well with the sum-field due to individual coils, eliminating the need for a fully self-consistent field calculation. A hollow permanent magnet could be used instead of the coil in Fig. 1, but this would sacrifice some of the flexibility to change the field strength.

The topology of the field is different above and below a characteristic value $B_{bias}=B_{thres}$ (Figure 2). For $B_{bias}<B_{thres}$ there are two quadrupole minima with different gradient on the symmetry axis x=0,y=0. Both of these can be used to load atoms using magneto-optical trapping techniques, but opposite circular polarizations would be required for each of them. As $B_{bias}$ is increased the minima move towards each other until they merge for $B_{bias}=B_{thres}$. At this point a single minimum exists with hexapole symmetry. In the single ring charge approximation the minimum is located where the z-derivative of Eq. (1) takes its maximum value. This is $z_{thres}=0.707$. When the ring charge on the other side of the dipole is taken into account $z_{thres}$ changes, but by less than 20% in most cases. This provides a good estimate for where the trapping lasers should cross to capture atoms from a hot background vapor. When $B_{bias}$ is increased further, the holding field assumes a ring topology. The radius of this ring grows while increasing $B_{bias}$ until it reaches the radius of the mu-metal shield. Further increasing $B_{bias}$ will increase the confinement (gradient in the radial direction), but hardly changes the location of the ring.

We emphasize that atoms can be *magneto-optically* trapped from the background vapor in a cell, both for $B_{bias}<B_{thres}$ (spherical cloud) but also $B_{bias}>B_{thres}$ (ring). Figure 2 shows a vector plot of the field in both situations. Previous experiments, to be described hereafter, have shown that magneto-optical trapping works well if the propagation direction and the magnetic field direction are within 15 degree of each other and if the correct (circular) polarization is applied. Magneto-optical trapping of a spherical cloud of atoms has been studied in detail [21] and serves as a departure point to achieve BEC. Specifically, using optical pumping into a dark state, a large initial density –enough to start evaporative cooling- can be created. This could well serve to create phase coherence before the cloud is mapped onto a ring.

To avoid Majorana "spin-flip" transitions to untrapped states of the magnetic potential an extra plugging field needs to be applied to remove the magnetic degeneracy. For a torus a current carrying wire along the z-axis can achieve this. When a thermal sample is mapped onto the ring this "piercing" wire may be added later using a motorized stage, as the spin flip rate is usually small. A more elaborate sequence of magnetic field manipulations is required when mapping a quantum degenerate cloud, to prevent loss of atoms.

3. Transformation of an atom cloud: experiment.

To demonstrate the possibility of continuously transforming a magnetic potential from a single well to a double well, while maintaining magnetic or magneto-optical trapping, we shall now describe an experiment using a straight guide with four mu-metal shields and a *horizontal* external bias field $B_{bias}$. (Fig. 3). Although this geometry is different from that in Fig. 1, we will use it to point out that a single minimum far away from a surface splits up closer to the surface where the finer structure of the magnetic field on that surface becomes apparent. It is also demonstrated that magneto-optical trapping can be maintained while the splitting occurs.

We consider the outer shields in Fig. 3 as a pair with equal but oppositely poled magnetization, and likewise for the inner pair. Without an external bias the field of the inner coils cancels that of the outer coils in exactly one point in the transverse x-z plane, creating a single waveguide in the y direction. With a horizontal bias field two minima exist, corresponding to two parallel magnetic guides. By either tuning the external bias field or the inner coil magnetization these minima can be made to merge or move apart, in a similar fashion as the ring magnet described earlier. Splitting two guides has been suggested for a similar system using parallel wires instead of coils [24].

In the geometry of Fig. 3 we use one set of "molasses" laser beams along the y-axis and another set at 45 degree angles with the x-y plane, in which a mirror is located (no $\lambda/4$ plate for this geometry). These angles are set by the requirement that the propagation vector of the laser light be parallel with the vector direction of the magnetic field, which has rotated by 45 degrees in comparison to the vector direction in Fig. 2. However, in both cases the magnetic field is quadrupolar in the relevant dimensions: the field strength grows linearly with distance from the center, and the magneto-optical trapping forces are identical.

Immediately after switching off the magneto-optical trapping beams we take an absorption image along the y-axis, i.e. the long axis of the atom guide(s). The results are shown in Fig. 4. Similarly to Fig. 2 a shallow magnetic minimum moves closer to the surface while increasing $B_{bias}$, and then splits laterally. The two minima of the double well are both filled with atoms at the same time using magneto-optical forces. Small asymmetries are observed which may be attributed to fabrication defects and alignment of the trapping laser beams. For large bias field the minima seem more separated in the experiment than in the theory, which is based on infinitely thin coils (thickness 0.5 mm in the experiment) and on the assumption that fields due to individual coils can be linearly superposed. This model breaks down for high field strengths due to effects of saturation and depolarization. In the numerical model we find that the gradient in a minimum scales as $B'_x = \alpha(B_{bias} - B_{thres})$ for $B_{bias} > B_{thres}$, with $\alpha = 17.8$ cm$^{-1}$ for the geometry and settings in the experiment.

4. Atom interferometry: effect of corrugations

Through the use of a ferromagnetic shield, the field due to individual wires is integrated. This boosts the field gradient and also averages out field corrugations due to anomalous current distributions inside the wires, a mechanism that was shown to cause fragmentation of Bose Einstein Condensates in atom chips [25]. Another source of field corrugations, perhaps the most significant one, is surface roughness in the material from which the magnetic field emerges. This surface roughness needs to be minimized by improving manufacturing technology, both in the case of atom chips and for the ferromagnetic coils presented here. To examine the effect of surface roughness for the ferromagnetic guide we may again adopt a monopole line-charge model $E_z = \frac{\lambda}{2\pi\varepsilon_0 z}$ at Z=0, using the equivalence of electric and magnetic dipole fields $\vec{B} \leftrightarrow \vec{E}$ (the contribution of the linear charge density λ at Z = -D can be neglected for Z<<D). For simplicity we assume a "slow" height variation dz of the coil at z=0, resulting in a variation d|B$_z$| at height z, so that $\frac{d|B_z|}{B} = \frac{|dz|}{z}$. Assuming that we can "polish" the surface to a roughness dz=0.5 μm, we find that the energy surface of the guide at a typical height z=100 μm changes 0.5% for various positions along the guide.

To examine the effect on reciprocal atom interferometry we have numerically solved the one-dimensional time-dependent Schrödinger equation $-\frac{\hbar^2}{2m}\frac{d^2\Psi}{ds^2} + V(s)\Psi = \frac{\hbar}{i}\frac{d\Psi}{dt}$ for propagation of a wavepacket in the presence of a randomly generated roughness potential given by $V(s) = A\sum_i^N \cos(i*\frac{s}{R} + \phi_i)$, where s is the distance measured along the ring guide and the spatial frequencies of the corrugated potential are assumed to have equal amplitude, but random phase $\phi_i \in \left|-\pi,\pi\right\rangle$, up to a maximum frequency limited by i=N. We have truncated at N=15 for our purpose, although the experimentally observed power spectrum of the surface corrugation may have a somewhat different shape. Gravity can be included in the first term of this expansion series. Cyclic boundary conditions have been correctly implemented by requiring continuity at the endpoints: $\Psi_{s=-\pi R} = \Psi_{s=+\pi R}$ and $\frac{d\Psi}{ds}|_{s=-\pi R} = \frac{d\Psi}{ds}|_{s=+\pi R}$. In the numerical implementation we have chosen units such that $\hbar = 1$, m=1 and 2πR=20. The one-dimensional model is a good approximation when the transverse motion in the guide is frozen and other vibrational states are energetically inaccessible; a more detailed analysis for >1 dimensions is given in [26].

The results of the simulation are plotted in Figure 5. We have taken an artifical initial wavepacket $\Psi_{s,t=0} = e^{-s^2/5} \times \cos(5s)$, which evolves into two counterpropagating components with average scaled velocity v=5. In reality atom diffraction using a optical standing wave could be used as a beamsplitting mechanism [19], in which case more components (diffraction orders) and larger periodicity are to be expected ($v = \pm 2nv_{recoil}$ with v$_{recoil}$=6 mm/s for rubidium, n=0,1,2…). Out of these we could then select the 1$^{st}$ and -1$^{st}$ diffraction order to examine the preservation of spatial coherence. It is observed that an interference pattern with visibility V~1 can be retrieved on both sides of the ring as long as the maximum height E$_{barrier}$ of the corrugation potential (of order: $\sqrt{N} \times A$) is smaller than half the kinetic energy. In the case of optical diffraction of rubidium (E$_{recoil}$=180 nK) we therefore can tolerate fractional magnetic field variations |dB$_{z=zmin}$|<E$_{recoil}$/μ$_B$=2.5 mG in the potential minimum of the guide. With a surface roughness dz/z=0.5% this leads to a maximum permissible field variation of 0.5 Gauss over the extension of the cloud, compatible with today's BEC experiments (where we have neglected the "plug" field that must be used to avoid Majorana transitions). Furthermore, in the regime where atoms hardly reflect off the corrugated potential ($E_{kin} \gg E_{barrier}$) and in the absence of rotation, fringes in the observed density pattern will only shift marginally after one full revolution (phase *difference*$\ll 2\pi$), making the interferometer suitable for gyroscopy.

It should be noted that reciprocal atom interferometry may be desirable with a thermal sample of atoms or even a single atom, using the fact that the rotational phase shift is independent of velocity ("white" interferometer). Although Bose Einstein Condensation has the capability of producing macroscopic phase coherence, it also raises the problem of phase "distortion" through collisions with thermal atoms, impeding sensor applications using the Sagnac effect. Rather than fighting the healing effect, one might make good use of it by constructing a rotation sensor based on the position of bright or dark solitons: shape preserving excitations of the condensate which can be dynamically stable in the 1D limit [22,23], and which are tolerant of certain levels of environmental noise. Another strategy to minimize the effects of corrugations is to use quantum-degenerate fermions, filling up the energy levels above the level of corrugations, after which coherent excitation is applied to the Fermi-surface. In any case an experimental study of quasi-1D phase coherence in a torus seems quite feasible: a linear array of $10^4$ Rb atoms laser cooled to the photon recoil temperature can be matched to a ring with 7.8 mm circumference.

5. Quantum Information

Consider a single atom stored in a ring. A Q-bit in the external degrees of freedom may be defined as the superposition of left hand circular (LHC) and right hand circular (RHC) momentum states: $|\Psi\rangle = \alpha|LHC\rangle + \beta|RHC\rangle$, with $\sqrt{\alpha^2 + \beta^2} = 1$. Diffraction in 1st and -1st orders, described before, would result in $|\Psi\rangle_{diffr} = \frac{1}{\sqrt{2}}\left(|-2\hbar k\rangle + e^{i\phi}|+2\hbar k\rangle\right)$, where the phase $\phi$ is set by the chosen diffraction mechanism. Similar to what has been realized in solid state physics [11], different ring waveguides may be coupled together by using two of the coils shown in Fig. 1, sharing the same vertical bias field $B_{bias}$. Figure 6 shows an equipotential plot of the energy surface $|B|=0.3B_{bias}$ for two coils with a diameter of 4 mm, separated by 1 mm in the junction. The system may be engineered such that a double well potential topological surface exists where the two guides approach each other. An additional "plugging" field needs to be applied to avoid Majorana transitions when the device is operated in the quantum degeneracy regime, in which case it becomes the equivalent of a Josephson junction. The tunneling rate can be adjusted with $B_{bias}$, but we find that the smallest potential barrier between the minima is not positioned on a straight line connecting the minima. This system can be mapped onto recent proposals for quantum computation with neutral atoms [27], with each ring storing a single Q-bit. A chain of rings forms a macroscopic realization of the 1D Ising model for an S = ½ spin system.

## 6. Conclusion

In summary we have presented a novel method to fill a toroidal magnetic guide with ultra cold atoms, using magneto-optical trapping or adiabatic transfer from a previously filled spherical/elliptical reservoir. Our recent observation of multiple caustics in the trajectories of atoms in ferromagnetic guides indicates that the magnetic field is smooth and well understood [28], although this has yet to tested in the limit of quantum degeneracy. Nevertheless, using a single, precisely manufactured, seamless and annealed ferromagnetic tube, our design may offer an attractive alternative to lithographically patterned devices.

## Acknowledgement

The author thanks Prof. Mara Prentiss for sponsorship and valuable discussions in the time that this work was completed.

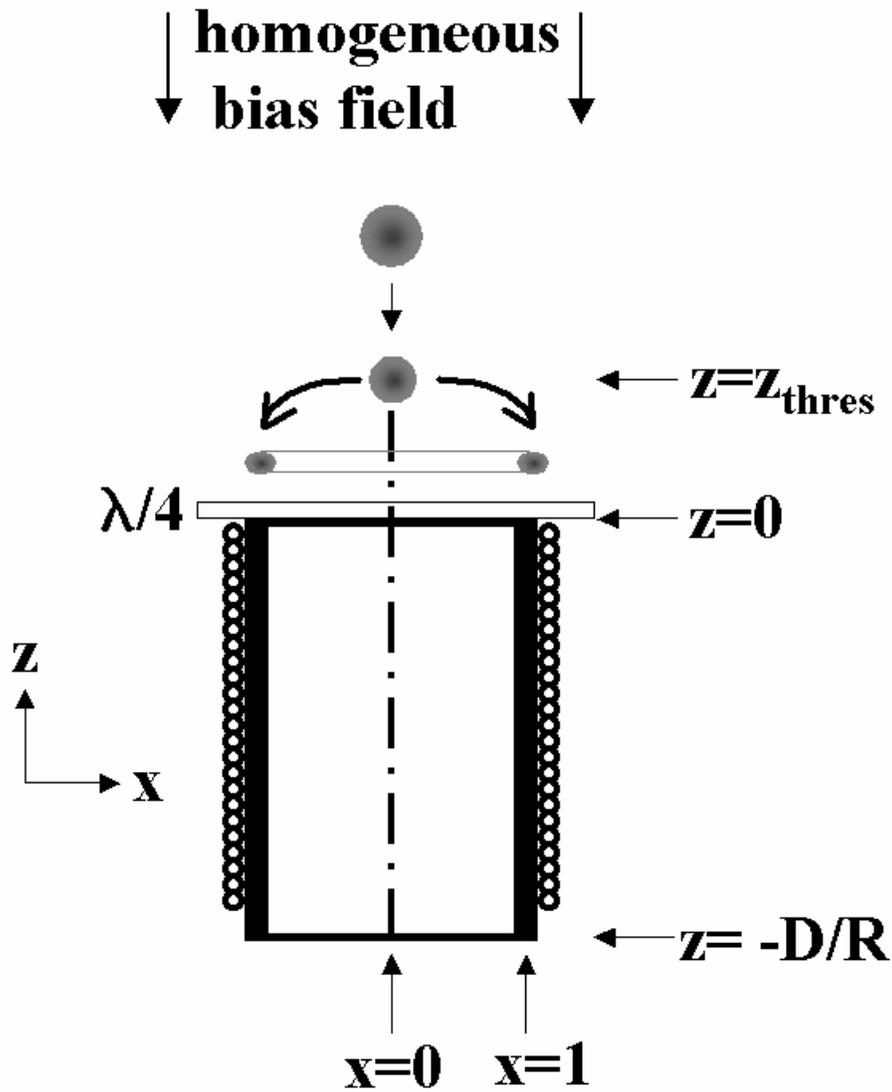

Figure 1: System used for creating toroidal magnetic trapping field: a mu-metal tube is wound with isolated copper wire. Combined with a homogeneous external field (as indicated) field minima exist above the tube which serve to contain an atom cloud. Increasing the bias field causes one of the minima to move towards the surface. The point-like minimum changes into a toroidal minimum at a critical height $z_{thres}$. A quarter-wave plate with a mirror coating on the back is necessary to reflect the laser beams that are used for magneto-optical trapping.

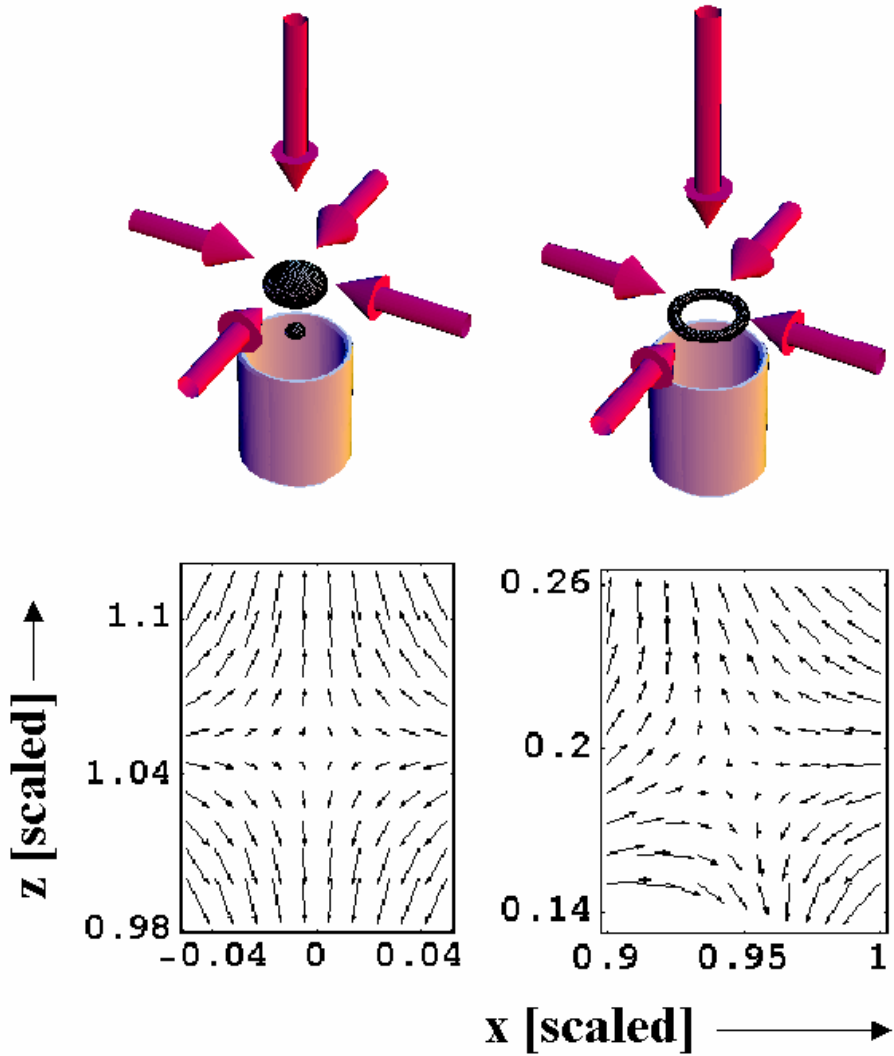

Figure 2: Top: geometry for magneto-optical and magnetic trapping of a cold atom cloud in a torus. Arrows indicate the laser beams for magneto-optical trapping. Top left: $B_{bias}=0.8B_{thres}$, two point-like minima with different gradient. Top right: $B_{bias}=4B_{thres}$, one torus-shaped minimum. The minima are indicated by an equipotential surface with constant |B|. Bottom: vector plots around the trap minima indicating the field direction.

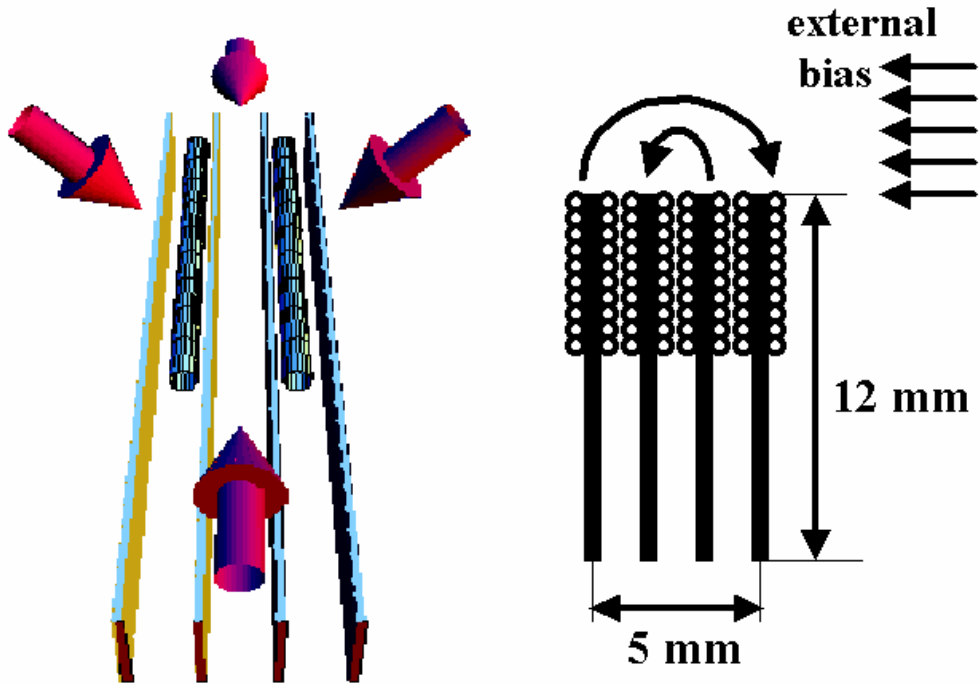

Figure 3: Setup for straight waveguide, as used in the experiment. Left: three-dimensional schematic with arrows indicating the trapping laser beams. A mirror (not shown) is mounted on top of the four mu-metal coils. Right: two-dimensional cross section of the guiding structure, using four mu-metal coils (in black, 10 windings each). Arrows indicate the magnetic field emerging from the coils and a homogeneous horizontal bias field. A combination of these fields results in line shaped field minima that can split and recombine.

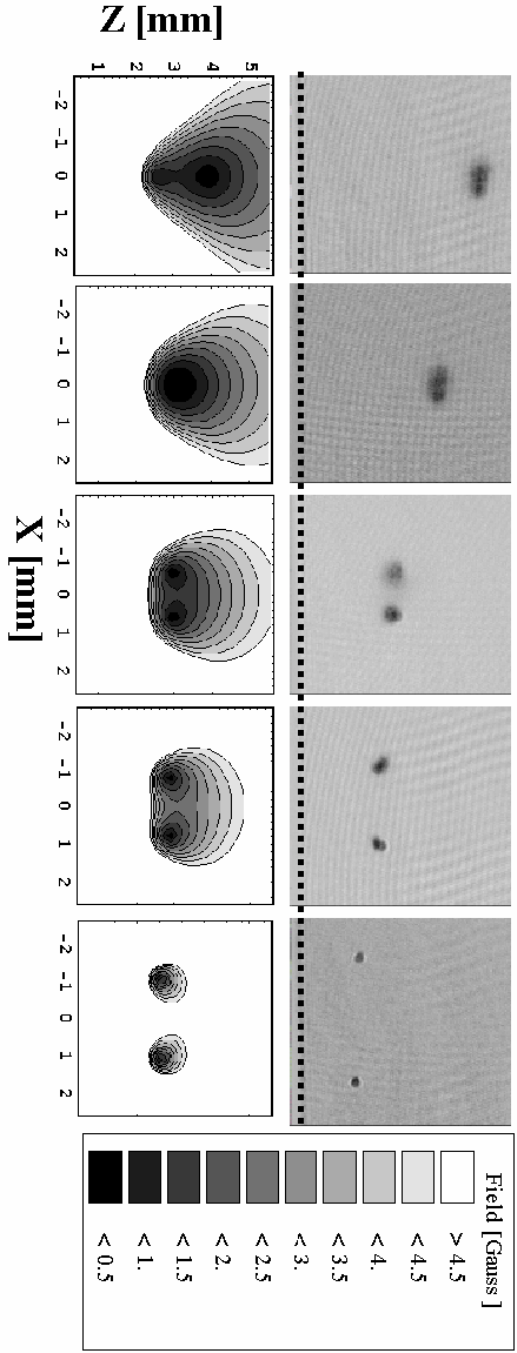

Figure 4: Top: Resonant absorption images (frame size: 4.9 mm x 4.9 mm) in the long direction of the atom cloud, immediately after the trap laser light is switched off, with horizontal bias field $B_{bias}$=5, 6, 7, 8 and 11 Gauss respectively. The four ferromagnetic coils carry a constant current of 1.0 A each. Dotted line corresponds to mirror surface. Bottom: equipotential contour plots, based on calculation which produces a gradient $B'_x$ = 14, 0, 33, 52, 101 G/cm in the respective minima. Scaling factor of coil field is adjusted to comply with $B_{thres}$=6G.

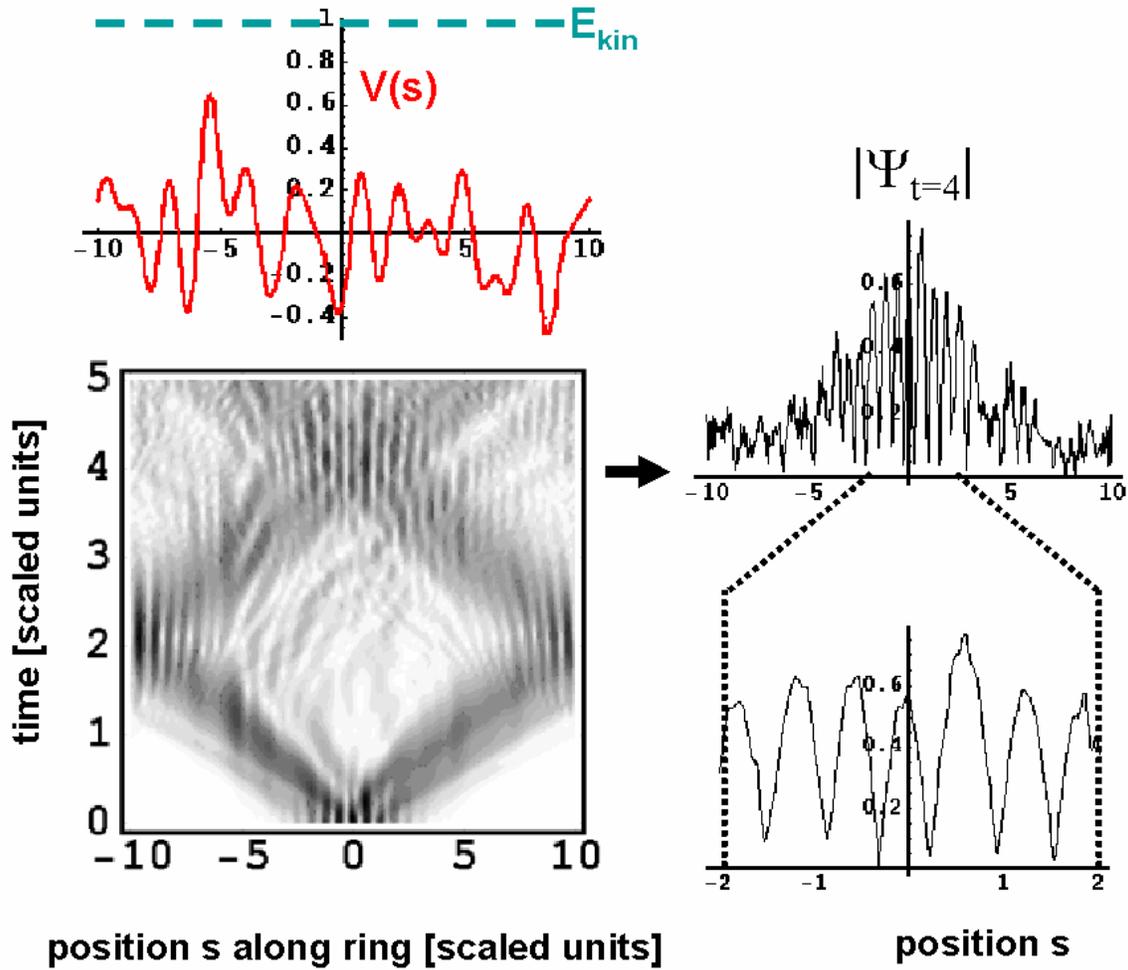

Figure 5: Left: Numerical solution of the time-dependent 1D Schrödinger equation in a ring geometry for an initial wavepacket $\Psi_{s,t=0} = e^{-s^2/5} \times \cos(5s)$. The randomly generated, corrugated potential V(s), shown at the top, is plotted in units of $mv^2/2$, where v=5 is the scaled mean velocity of the counterpropagating wavepackets. Inverted grayscale is used to represent density of atoms, i.e. white represents no atoms, black represents many atoms. Right: corresponding wavefunction after one roundtrip at t=4 (scaled units). The center fringe has hardly shifted and contrast remains of order unity.

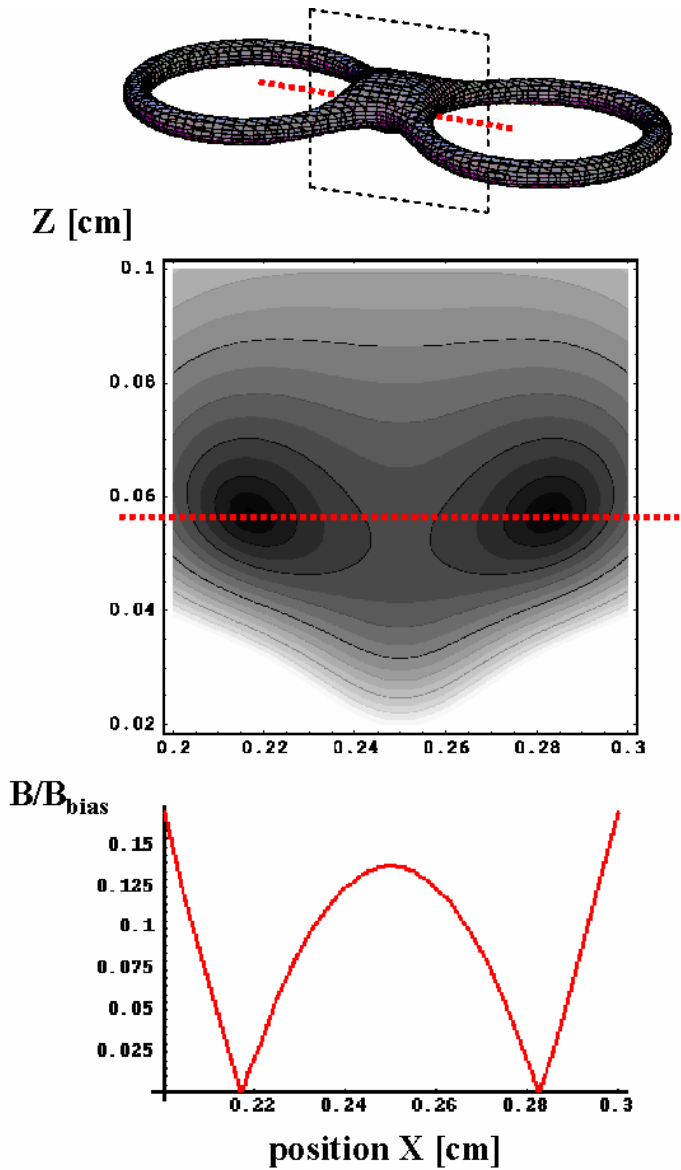

Figure 6: Top: Three-dimensional plot of the equipotential surface $|B|=0.38\ B_{bias}$, for two ferromagnetic coils placed next to each other (R=0.2 cm, distance between the symmetry axes of each coil: L=0.5 cm). Middle and bottom: corresponding two-dimensional and one-dimensional plot of the magnetic field in the junction, as indicated at the top. A typical gradient of 100 G/cm outside of the junction is produced using $B_{bias}$=6.8 G and corresponds to a barrier height of 0.85 G for this geometry.